\documentclass[twocolumn,showpacs,preprintnumbers,amsmath,amssymb]{revtex4}
\usepackage{amssymb}
\usepackage{txfonts}

\usepackage{graphicx}
\usepackage{dcolumn}
\usepackage{bm}
\usepackage{sidecap}

\begin{document}
\title{Security bound of continuous-variable quantum key distribution\\
 with noisy coherent states and channel}

\author{Yong Shen}
\author{Jian Yang}
\author{Hong Guo }\thanks{Author to whom correspondence should be addressed. E-mail: hongguo@
pku.edu.cn, Phone: +86-10-6275-7035, Fax: +86-10-6275-3208.}
\affiliation{CREAM Group, State Key Laboratory of  Advanced Optical
Communication Systems and Networks (Peking University) and Institute
of Quantum Electronics, School of Electronics Engineering and
Computer Science, Peking University, Beijing 100871, PR China}

\begin{abstract}
Security of a continuous-variable quantum key distribution protocol
based on noisy coherent states and channel is analyzed. Assuming the
noise of coherent states is induced by Fred, a neutral party
relative to others, we prove that the prepare and measurement scheme
and entanglement-based scheme are equivalent. Then, we show that
this protocol is secure against Gaussian collective attacks even if
the channel is lossy and noisy, and further, a lower bound to the
secure key rate is derived.
\end{abstract}

\maketitle

The principles of quantum mechanics make it possible to distribute
physically secure keys between two distant parties
\cite{BB84,Gisin02}. In particular, continuous-variable quantum key
distribution (CV-QKD) has made remarkable achievements during the
past few years. Several CV-QKD schemes based on coherent states
combined with homodyne or heterodyne detection have been proposed
\cite{1,2} and experimentally demonstrated \cite{3}--\cite{18}, and
the unconditional security of CV-QKD with ideal optical source has
also been systematically studied \cite{5,6,4,19}. However, there
still have difficulties to calculate the secure key rate of the
CV-QKD protocols with noisy coherent states, because of the absence
of a good model to characterize the source imperfections. For
instance, it is not a good idea to ascribe the source imperfection
to the channel excess noise directly, which implies that Eve is able
to control such noise. In this case, Eve actually knows that Alice
is using a P\&M scheme instead of an E-B one, and the two schemes
are no longer equivalent. Further more, from a practical viewpoint,
it is suspicious to assume that Eve is able to control the noise
inside Alice's side. The same topic was discussed recently \cite{7},
in which the noise of source and modulation is controlled neither by
legitimate users nor eavesdroppers. However, under this assumption,
Eve can not purify the states of Alice and Bob, which implies that
the previous proof on the optimality of Gaussian attacks \cite{11}
is not available. As a result, the key rate derived in \cite{7} may
be insecure under collective attack.

To solve these problems, in this paper, we propose a new way to
characterize the source and modulation noise and derive a security
bound of the CV-QKD protocol through a lossy and noisy channel. As
we know, in a standard prepare and measurement (P$ \&$M) scheme,
Alice generates two Gaussian random numbers, $Q_{A}$ and $P_{A}$,
with mean values $0$ and variances $V_{A}$, to prepare
$|Q_{A}+iP_{A}\rangle$ by modulating an initial coherent state.
Then, she sends this state to Bob through a quantum channel
characterized by transmittance $T$ and excess noise
$\varepsilon_{C}$. Receiving the state, Bob randomly chooses one
quadrature to measure and informs Alice which observable he
measured. Then, Alice and Bob should share two correlated Gaussian
variables, from which they can extract a private binary key with
standard reconciliation and privacy amplification process. However,
in a realistic case, the optical source and modulators are imperfect
and inevitably induces extra excess noise (denoted by two Gaussian
random numbers $\Delta Q_{A}$ and $\Delta P_{A}$) to the coherent
state. As a result, Alice actually prepares a state of
$|(Q_{A}+\Delta Q_{A})+i(P_{A}+\Delta P_{A})\rangle$, where $\Delta
Q_{A}$ and $\Delta P_{A}$, independent with $Q_{A}$ and $P_{A}$,
have mean values $0$ and variances $\varepsilon_{0}$ \cite{7}.
\begin{figure}
\centering
\includegraphics[width=8.5 cm]{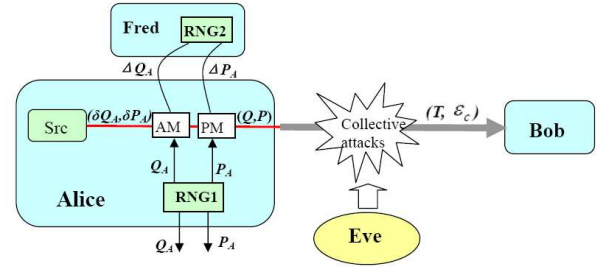}
\caption{(color online) The P$\&$M scheme. A random number generator
(RNG) gives two values $Q_A$ and $P_A$. The coherent state is
generated in source (Src) with shot noise ($\delta Q_A,\delta P_A$),
and then, its central position in phase space is displaced by
Gaussian modulations with $(Q_A+\Delta Q_A,P_A+\Delta P_A)$. The
noise $\Delta Q_A$ and $\Delta P_A$ is assumed to be individually
induced by a neutral party, Fred.}\label{pic1}
\end{figure}
In a trusted-source model, neither Eve nor Alice can control this
excess noise. So, it is convenient to phenomenologically assume that
the noise of coherent state is not induced by the imperfection of
devices but by a neutral party, Fred, who generates two random
numbers, $\Delta Q_{A}$ and $\Delta P_{A}$, and then introduces
corresponding extra noise to the coherent state after an ideal
modulation (See Fig. 1). In this case, the state sent to Bob is
denoted by quadratures ($Q$,$P$), which satisfy
\begin{equation}
\begin{array}{l}
 Q = Q_A  + \delta Q_A  + \Delta Q_A , \\
 P = P_A  + \delta P_A  + \Delta P_A , \\
 \end{array}
\end{equation}
where $\delta Q_A$ and $\delta P_A$ are originated from shot noise,
and satisfy $\left\langle {\delta Q_A ^2 } \right\rangle =
\left\langle {\delta P_A ^2 } \right\rangle  = 1$, in shot-noise
units. The conditional variance $V_{Q|Q_A}$ is \cite{10}
\begin{equation}
V_{Q|Q_A }  = \left\langle {Q^2 } \right\rangle  -
\frac{{\left\langle {QQ_A } \right\rangle ^2 }}{{\left\langle {Q_A
^2 } \right\rangle }} = (1 + \varepsilon _0).
\end{equation}

\begin{figure}
\centering
\includegraphics[width=8.5 cm]{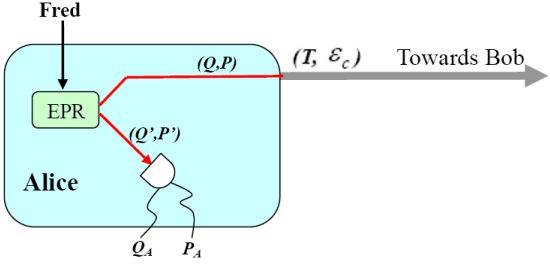}
\caption{(color online) The E--B scheme. Fred prepares $\rho_{AB}$
for Alice. In order to prepare a coherent state for Bob, Alice can
measure both quadrature of her beam with a balanced heterodyne
detector.}\label{pic2}
\end{figure}

For the convenience of theoretical analysis, here we propose an
entanglement based (E-B) scheme and will show it is equivalent to
the P$\&$M scheme. As shown in Fig. \ref{pic2}, Fred prepares a pair
of EPR beams $\rho_{AB}$, and holds its purification. The global
pure state shared by Alice, Bob and Fred is denoted by $\left| {\psi
_{ABF} } \right\rangle$, and $\rho_{AB}$ is assumed to be a Gaussian
state. Quadratures $(Q,P)$ denote the state sent to Bob, and
$(Q',P')$ denote the state kept by Alice. Here, we assume $(Q,P)$
and $(Q',P')$ satisfy
\begin{equation}\label{1}
 \left\langle {Q^2 } \right\rangle  = \left\langle {P^2 } \right\rangle= (V+\varepsilon _0),\quad \left\langle {{Q'}^2 }
 \right\rangle  = \left\langle {{P'}^2 } \right\rangle= V ,
\end{equation}
where $V=V_A+1$. According to \cite{10}, we obtain
\begin{equation}\label{2}
 |\left\langle{QQ'}\right\rangle|^2 \le\left\langle{Q^2}\right\rangle\left\langle{{Q'}^2}\right\rangle-\frac{\left\langle
 {{Q'}^2}\right\rangle}{\left\langle{P^2}\right\rangle}= (V+\varepsilon _0)V-\frac{V}{V+\varepsilon
_0}.
\end{equation}
We assume that
\begin{equation}
\left\langle{QQ'}\right\rangle=\sqrt{V^2-1},
\end{equation}
which absolutely satisfies Eq. (\ref{2}). Similarly, we assume that
\begin{equation}
\left\langle{PP'}\right\rangle=-\sqrt{V^2-1}.
\end{equation}
In E-B scheme, Alice makes a balanced heterodyne detection on $Q'$
and $P'$ simultaneously. Denoting the values of $Q'$ and $P'$
measurements as $Q'_A$ and $P'_A$, we have \cite{10}
\begin{equation}
Q' = {Q'}_A  + \delta Q'_A ,\quad P' = {P'}_A  + \delta P'_A,
\end{equation}
where $\left\langle {{\delta Q_A'} ^2 } \right\rangle  =
\left\langle {{\delta P_A'} ^2 } \right\rangle  = 1$. Alice's best
estimate of $(Q,P)$ is denoted by $(Q_A,P_A)$, which satisfy
\cite{10}
\begin{equation}
Q_A  = \sqrt {\frac{{V - 1}}{{V + 1}}} Q'_A ,\quad P_A  = - \sqrt
{\frac{{V - 1}}{{V + 1}}} P'_A .
\end{equation}
We can easily get
\begin{equation}
\begin{array}{l}
 \left\langle {Q_A ^2 } \right\rangle  = \left\langle {P_A ^2 } \right\rangle  = \left( {V - 1} \right)  =
 V_A,\\[6pt]
 V_{Q|Q_A }  = V_{P|P_A }  =(1+\varepsilon _0) ,\\[6pt]
 \end{array}
\end{equation}
which shows the same result as in P$\&$M scheme. Let us suppose the
EPR source and the measuring apparatus of Alice are hidden in a
black box. The only outputs of this black box are the values of
$\varepsilon _0$, $Q_A$ and $P_A$, and the beam $(Q,P)$. For Eve,
this black box is indistinguishable from the equivalent black box,
sketched in the P$\&$M scheme. So, we can calculate the secure key
rate in the E-B scheme. When Eve makes collective attacks, she
interacts individually with each quantum state in the same way.
Accordingly, the global state $\left| {\psi _{ABF} } \right\rangle$
becomes $\left| {\psi _{ABEF} } \right\rangle$. After Alice and
Bob's measurements, Alice, Bob and Eve should share correlated
Classical-Classical-Quantum information (CCQ correlations)
\cite{12}, and the security key rate can be calculated by \cite{8}
\begin{equation}
\begin{array}{l}
 K_D  = I(a:b) - \chi (a:E), \\[6pt]
 K_R  = I(a:b) - \chi (b:E), \\[6pt]
 \end{array}
\end{equation}
where $K_{D}$ and $K_{R}$ correspond to the direct and reverse
reconciliation, respectively. $I(a:b)$ is the Shannon mutual
information between Alice and Bob. $\chi (a:E) = S(\rho _E ) - \int
{P(a)S(\rho _E^a )} {\rm{d}}a$ and $\chi (b:E) = S(\rho _E ) - \int
{P(b)S(\rho _E^b )} {\rm{d}}b$ are Holevo bounds \cite{9}, which put
an upper limit on how much information can be contained in a quantum
system. $S(\rho_E)$ is the von Neumann entropy of Eve's state
$\rho_E$, and $S(\rho_E^a)$ [$S(\rho_E^b)$] is the von Neumann
conditional entropy of $\rho_E$ while knowing the measurement value
$a$ ($b$) of Alice (Bob). However, since $\rho_{ABE}$ is not a pure
state, and the optimality of Gaussian attacks has not been proved,
we can not calculate a minimum $K_D$ or $K_R$ directly. Fortunately,
if we suppose the state of Fred can be controlled by Eve, and $|\psi
_{ABEF}\rangle$ is a pure state, we can derive a lower bound of
$K_D$ or $K_R$, denoted by $\tilde K_D$ and $\tilde K_R$, where
\begin{equation}
\begin{array}{l}
 \tilde K_D  = I(a:b) - \chi (a:EF), \\
 \tilde K_R  = I(b:a) - \chi (b:EF), \\
 \end{array}
\end{equation}
where $\chi (a:EF)$ and $\chi (b:EF)$ are the Holevo bounds between
Alice and the eavesdroppers. It can be proved that \cite{14}
\begin{equation}
 \quad\chi (a:EF) - \chi (a:E)\ge 0.
\end{equation}
Since $\rho_{ABEF}$ is a pure state, it has been proved that $\tilde
K_D$ and $\tilde K_R$ get their minimal values when $\rho_{AB}$ is a
Gaussian state \cite{11}. The Shannon mutual information is given by
$I(a:b) = {1 \mathord{\left/
 {\vphantom {1 2}} \right.
 \kern-\nulldelimiterspace} 2}\log _2 ({{V_b } \mathord{\left/
 {\vphantom {{V_b } {V_{b|a} }}} \right.
 \kern-\nulldelimiterspace} {V_{b|a} }})$, where $V_b$ is the
variance of the random variable of Bob, and $V_{b|a}$ is the
conditional variance of the random variable of Bob while knowing
Alice's measurement. Here, we can only calculate the case of $Q$
quadrature for the symmetry in $Q$ and $P$. Let $Q_{B'}$ be the
result of Bob's measurement on the $Q$ quadrature and $Q_A$ be
Alice's best estimate of $Q_{B'}$. We obtain
\begin{equation}
\begin{array}{l}
 V_b  = \left\langle {Q_{B'}^2 } \right\rangle  = T(V + \chi ), \\[6pt]
 V_{b|a}  = \left\langle {Q_{B'}^2 } \right\rangle  - \frac{\displaystyle{\left\langle {Q_A Q_{B'} } \right\rangle }^2}
 {\displaystyle{\left\langle {Q_A^2 } \right\rangle }} = T(1 + \chi ),
 \\[6pt]
 \end{array}
\end{equation}
where $\chi  = {{(1 - T)} \mathord{\left/
 {\vphantom {{(1 - T)} T}} \right.
 \kern-\nulldelimiterspace} T} + \varepsilon _0+\varepsilon _c$, and $\varepsilon_{c}$ denotes the excess noise in the channel. Hence
\begin{equation}
I(a:b) = \frac{1}{2}\log _2 \left(\frac{{V + \chi }}{{1 + \chi
}}\right).
\end{equation}
Since $\rho_{ABEF}$ is pure, we have $S(\rho_{EF})$=$S(\rho_{AB})$.
So,
\begin{equation}
\begin{array}{l}
 \chi (a:EF) = S(\rho _{EF} ) - \int {P(a)S(\rho _{EF}^a )} {\rm{d}}a \\[6pt]
 \qquad\quad\;\;\; = S(\rho _{AB} ) - \int {P(a)S(\rho _{AB}^a )} {\rm{d}}a .\\[6pt]
 \end{array}
\end{equation}

\begin{figure}
\centering
\includegraphics[width=8.5 cm, height=6.5 cm]{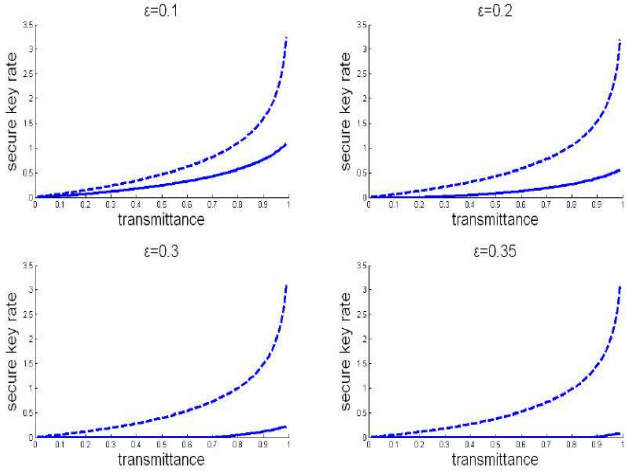}
\caption{(color online) The lower bound and the secure key rate as a
function of the transmittance of the channel; solid line--$\tilde
K_R^{{\rm{hom}}}$, dash line--$K_R^{{\rm{hom}}}$.}\label{pic3}
\end{figure}

For a two-mode Gaussian state $\rho_{AB}$, we can calculate its von
Neumann entropy with its covariance matrix $\gamma_{AB}$ \cite{13},
which can be expressed as
\begin{equation}
\gamma _{AB}  = \left[ {\begin{array}{*{20}c}
   A & C  \\
   {C^T } & B  \\
\end{array}} \right],
\end{equation}
where $A$, $B$ and $C$ are $2 \times 2$ blocks. Let $\Delta(\gamma
_{AB} )  = \det A + \det B + 2\det C$, and
\begin{equation}
s_{1,2}  = \sqrt {\frac{{\Delta(\gamma _{AB} )  \pm \sqrt {\Delta
^2(\gamma _{AB} ) - 4\det (\gamma _{AB} }) }}{2}}.
\end{equation}
Then the von Neumann entropy of $\rho_{AB}$ is \cite{13}
\begin{equation}
S(\rho _{AB} ) = g(s_1 ) + g(s_2 ),
\end{equation}
where
\begin{equation}
g(x) = \frac{{x + 1}}{2}\log _2 \left( {\frac{{x + 1}}{2}} \right) -
\frac{{x - 1}}{2}\log _2 \left( {\frac{{x - 1}}{2}} \right).
\end{equation}
In the case where Alice prepares coherent states and Bob makes
homodyne detection, it is easy to see that the covariance matrix of
$\rho_{AB}$ is
\begin{equation}
\gamma _{AB}  = \left[ {\begin{array}{*{20}c}
   {x\mathbb{I} } & {z \sigma }  \\
   {z \sigma } & {y\mathbb{I} }  \\
\end{array}} \right],
\end{equation}
where $x = V$, $y = T(V + \chi )$, $z = \sqrt {T(V^2  - 1)}$,
$\mathbb{I}=\rm{diag}(1,1)$ and $\sigma=\rm{diag}(1,-1)$. So
\begin{equation}
\begin{array}{l}
 \det (\gamma _{AB} ) = (T + T\chi V)^2 , \\[6pt]
 \Delta (\gamma _{AB} ) = V^2  - 2T(V^2  - 1) + (TV + T\chi )^2 . \\[6pt]
 \end{array}
\end{equation}
When Alice makes her measurement, she splits her mode into two beams
by a 50:50 beam splitter, and  measures the $Q$ and $P$ quadratures
of two beams respectively. If Bob chooses a quadrature to measure,
say, the $Q$ quadrature, then Alice discards the result of the
measurement on the $P$ quadrature. It is easy to see that the two
mode state $\rho_{AB}^a$ is still a Gaussian state. The covariance
matrix of this state is
\begin{widetext}
\begin{equation}
\gamma _{AB}^a  = \left[ {\begin{array}{*{20}c}
   {\displaystyle\frac{{2V}}{{V + 1}}} & 0 & {\sqrt {\displaystyle\frac{{2T(V - 1)}}{{V + 1}}} } & 0  \\
   0 & {\displaystyle\frac{{V + 1}}{2}} & 0 & { - \sqrt {\displaystyle\frac{{T(V^2  - 1)}}{2}} }  \\
   {\sqrt{\displaystyle\frac{{2T(V - 1)}}{{V + 1}}} } & 0 & {T(1 + \chi )} & 0  \\
   0 & { - \sqrt{\displaystyle\frac{{T(V^2  - 1)}}{2}} } & 0 & {T(V + \chi )}  \\
\end{array}} \right],
\end{equation}
\end{widetext}
and
\begin{equation}
\begin{array}{l}
 \det (\gamma _{AB}^a ) = (T + T\chi )(T + T\chi V), \\[6pt]
 \Delta (\gamma _{AB}^a ) = 2T + T^2\chi (1 + \chi ) + \left[ {T^2 \chi  + (1 - T)^2 } \right]V. \\[6pt]
 \end{array}
\end{equation}
It can be found that the von Neumann conditional entropy
$S(\rho_{AB}^a)$ does not depend on Alice's measurement $a$. So the
Holevo bound is simply equal to
\begin{equation}
\chi (a:EF) = S(\rho _{AB} ) - S(\rho _{AB}^a ).
\end{equation}
At the high modulation limit $(V \gg 1/(1-T), 1/T)$, the lower bound
of the secure key rate for the direct reconciliation is
\begin{equation}
\begin{array}{l}
 \tilde K_D^{{\rm{hom}}}  = {\displaystyle\frac{1}{2}}\log _2 \left[\displaystyle\frac{{T^2 \chi  + (1 - T)^2 } }{{1 + \chi
 }}\right]
 - \log _2 (1 - T)  \\[10pt]
 \qquad\;\, - g\left( \displaystyle{\frac{{T\chi }}{{1 - T}}}  \right)+  g\left[ \displaystyle{\frac{\sqrt{(1+
 \chi )\chi}T  }{\sqrt{T^2 \chi    +(1 - T)^2 }}} \right]. \\[10pt]
 \end{array}
\end{equation}
Similarly, we can derive $\tilde K_R^{{\rm{hom}}}$ in the high
modulation limit,
\begin{equation}
\tilde K_R^{{\rm{hom}}}  = \frac{1}{2}\log _2 \left(\frac{\chi }{{1
+ \chi }}\right) - \log _2 (1 - T) - g\left( {\frac{{T\chi }}{{1 -
T}}} \right).
\end{equation}
The limiting value of $\varepsilon_0$ is shown as
\begin{equation}
\varepsilon_0  = \frac{1}{2}\left( {\sqrt {1 +
\frac{{16}}{{{\rm{e}}^2 }}}  - 1} \right) \approx 0.39.
\end{equation}
We will give a demonstration on how much  the security bound is
lower than the key rate derived in \cite{7}. For comparing with the
result in \cite{7}, we just consider the effect of the noisy
coherent states, and any excess noise in the channel is now assumed
to be negligible. When the channel is noiseless, the key rate
derived in \cite{7} is
\begin{equation}
K_R^{{\rm{hom}}}  =  - \frac{1}{2}\log _2 \left[\left( {\frac{T}{{V
+ \varepsilon _0 }} + 1 - T} \right)T\left( {1+\chi} \right)\right],
\end{equation}
and in the high modulation limit it becomes
\begin{equation}
K_R^{{\rm{hom}}}  = \frac{1}{2}\log _2 \left[\frac{1}{{T(1 -
T)(1+\chi )}}\right].
\end{equation}
There is no limiting value of $\varepsilon_0$ in this case. If
$\varepsilon_0 =0$, $\tilde K_R^{{\rm{hom}}}$ and $K_R^{{\rm{hom}}}$
are the same, otherwise $\tilde K_R^{{\rm{hom}}}$ is obviously
smaller than $K_R^{{\rm{hom}}}$, as shown in Fig. \ref{pic3}.
The exact secure key rate is just between the solid line and the
dash line. So we can estimate how much information is lost when we
use the lower bound. However, if the noise of the source is too
large, the security bound will be too low and the gap between the
solid line and dash line will be too large. In this case, it is hard
to estimate how much information we lose. This problem can be solved
if we can effectively purify the coherent state and reduce the
excess noise \cite{7}.

In conclusion, we study the security of a continuous-variable
quantum key distribution protocol with noisy coherent states sent
through a lossy and noisy channel. Though the protocol discussed is
a P$\&$M scheme, it is proved to be equivalent to an E-B scheme when
we assume the noise in coherent states is induced by Fred, a neutral
party, who does not give Eve any information. Though the state
shared by Alice, Bob and Eve is not a pure state, and the secure key
rate can not be calculated directly, we can derive the maximal
mutual information between Eve and Alice/ Bob, if she is able to
acquire extra information from Fred. So, we actually derive a lower
bound to the secure key rate. We also give a demonstration that when
the channel is lossy but noiseless, the lower bound becomes lower
than the key rate derived in \cite{7} with the increase of the
excess noise in the coherent states. We can also purify the noisy
coherent states to get a better estimation on the information lost.

The authors thank Xiang Peng and Bingjie Xu for fruitful
discussions. The work is supported by the Key Project of National
Natural Science Foundation of China (Grant No. 60837004).

\end{document}